# Excitonic properties of strained wurtzite and zinc-blende GaN/Al$_x$Ga$_{1–x}$N quantum dots

Vladimir A. Fonoberov[*] and Alexander A. Balandin

*Nano-Device Laboratory, Department of Electrical Engineering*
*University of California–Riverside, Riverside, California 92521*

We investigate exciton states theoretically in strained GaN/AlN quantum dots with wurtzite (WZ) and zinc-blende (ZB) crystal structures, as well as strained WZ GaN/AlGaN quantum dots. We show that the strain field significantly modifies the conduction and valence band edges of GaN quantum dots. The piezoelectric field is found to govern excitonic properties of WZ GaN/AlN quantum dots, while it has a smaller effect on WZ GaN/AlGaN, and very little effect on ZB GaN/AlN quantum dots. As a result, the exciton ground state energy in WZ GaN/AlN quantum dots, with heights larger than 3 nm, exhibits a red shift with respect to the bulk WZ GaN energy gap. The radiative decay time of the red-shifted transitions is large and increases almost exponentially from 6.6 ns for quantum dots with height 3 nm to 1100 ns for the quantum dots with height 4.5 nm. In WZ GaN/AlGaN quantum dots, both the radiative decay time and its increase with quantum dot height are smaller than those in WZ GaN/AlN quantum dots. On the other hand, the radiative decay time in ZB GaN/AlN quantum dots is of the order of 0.3 ns, and is almost independent of the quantum dot height. Our results are in good agreement with available experimental data and can be used to optimize GaN quantum dot parameters for proposed optoelectronic applications.

## I.   INTRODUCTION

Recently, GaN quantum dots (QDs) have attracted significant attention as promising candidates for application in optical, optoelectronic, and electronic devices. Progress in GaN technology has led to many reports on fabrication and characterization of different kinds of GaN QDs [1–8]. Molecular beam epitaxial growth in the Stranski-Krastanov mode of wurtzite (WZ) GaN/AlN [1, 2] and GaN/Al$_x$Ga$_{1-x}$N [3, 4] QDs has been reported. Other types of WZ GaN QDs have been fabricated by pulsed laser ablation of pure Ga metal in flowing N$_2$ gas [5], and by sequential ion implantation of Ga$^+$ and N$^+$ ions into dielectrics [6]. More recently, self-organized growth of zinc-blende (ZB) GaN/AlN QDs has been reported [7, 8].

Despite the large number of reports on the fabrication and optical characterization of WZ GaN/AlN and GaN/Al$_x$Ga$_{1-x}$N as well as ZB GaN/AlN QDs, there have been a small number of theoretical investigations of electronic states and excitonic properties of GaN QDs [9, 10]. Electronic states in WZ GaN/AlN QDs have been calculated in Ref. [9] using the plane wave expansion method. In addition to the restrictions imposed by any plane wave expansion method, such as the consideration of only 3D-periodic structures of coupled QDs and the requirement of a large number of plane waves for QDs with sharp boundaries, the model of Ref. [9] assumes equal elastic as well as dielectric constants for both the QD material and the matrix. Ref. [10] briefly describes a calculation of excitonic properties of several specific types of GaN QDs.

In this paper, we present a theoretical model and numerical approach that allows one to accurately calculate excitonic and optical properties of strained GaN/Al$_x$Ga$_{1-x}$N QDs with WZ and ZB crystal structure. Using a combination of finite difference and finite element methods we accurately determine strain, piezoelectric, and Coulomb fields as well as electron and hole states in WZ GaN/AlN and GaN/Al$_x$Ga$_{1-x}$N as well as ZB GaN/AlN QDs. We take into account the difference in the elastic and dielectric constants for the QD and matrix (barrier) materials. We investigate in detail the properties of single GaN QDs of different shapes, such as a truncated hexagonal pyramid on a wetting layer for WZ GaN/AlN QDs [see Fig. 1(a)], a disk for WZ GaN/Al$_x$Ga$_{1-x}$N QDs, and a truncated square pyramid on a wetting layer for ZB GaN/AlN QDs [see Fig. 1(b)]. Our model allows direct comparison of excitonic properties of different types of GaN QDs with reported experimental data, as well as analysis of the functional dependence of these properties on QD size.

---

[*] Author to whom correspondence should be addressed; electronic address: vladimir@ee.ucr.edu

The paper is organized as follows. Sections II–VI represent the theory for calculation of excitonic properties of strained QD heterostructures with either WZ or ZB crystal structure. Calculation of strain and piezoelectric fields is described in Sections II and III, correspondingly. Section IV outlines the theory of electron and hole states in strained QD heterostructures. The Coulomb potential energy in QD heterostructures is given in Section V. Section VI demonstrates a method to calculate exciton states, oscillator strengths and radiative decay times. Results of the numerical calculations for different GaN QDs are given in Section VII. Conclusions are presented in Section VIII.

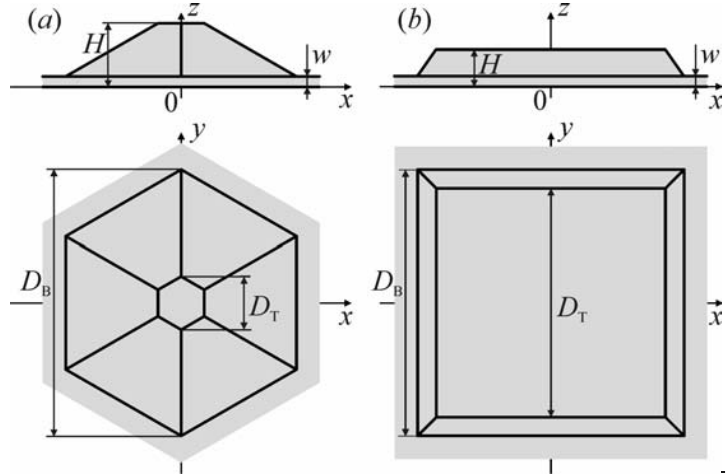

FIG. 1. Shapes of WZ GaN/AlN (a) and ZB GaN/AlN (b) QDs.

## II. STRAIN FIELD IN QUANTUM-DOT HETEROSTRUCTURES

The lattice constants in semiconductor heterostructures vary with coordinates. This fact leads to the appearance of the elastic energy [11]

$$F_{\text{elastic}} = \int_V d\mathbf{r} \sum_{ijlm} \frac{1}{2} \lambda_{ijlm}(\mathbf{r}) \varepsilon_{ij}(\mathbf{r}) \varepsilon_{lm}(\mathbf{r}), \quad (1)$$

where $\varepsilon_{ij}$ is the strain tensor, $\lambda_{ijlm}$ is the tensor of elastic moduli, $V$ is the total volume of the system, and $ijlm$ run over the spatial coordinates $x$, $y$, and $z$. To account for the lattice mismatch, the strain tensor $\varepsilon_{ij}$ is represented as [12]

$$\varepsilon_{ij}(\mathbf{r}) = \varepsilon_{ij}^{(\mathbf{u})}(\mathbf{r}) - \varepsilon_{ij}^{(0)}(\mathbf{r}), \quad (2)$$

where $\varepsilon_{ij}^{(0)}$ is the tensor of local intrinsic strain and $\varepsilon_{ij}^{(\mathbf{u})}$ is the local strain tensor defined by the displacement vector **u** as follows,

$$\varepsilon_{ij}^{(\mathbf{u})}(\mathbf{r}) = \frac{1}{2}\left(\frac{\partial u_i(\mathbf{r})}{\partial r_j} + \frac{\partial u_j(\mathbf{r})}{\partial r_i}\right). \tag{3}$$

To calculate the strain field [Eqs. (2), (3)] one has to find the displacement vector $\mathbf{u}(\mathbf{r})$ at each point of the system. This can be achieved by imposing boundary conditions for $\mathbf{u}(\mathbf{r}_\infty)$ at the endpoints $\mathbf{r}_\infty$ of the system and minimizing the elastic energy (1) with respect to $\mathbf{u}(\mathbf{r})$.

### A. Zinc-blende quantum dots

In crystals with ZB symmetry, there are only three linearly independent elastic constants: $\lambda_{xxxx} = C_{11}$, $\lambda_{xxyy} = C_{12}$, and $\lambda_{xyxy} = C_{44}$. Thus, the elastic energy (1) can be written as

$$F_{\text{elastic}} = \frac{1}{2}\int_V d\mathbf{r}\left(C_{11}(\varepsilon_{xx}^2 + \varepsilon_{yy}^2 + \varepsilon_{zz}^2) + 2C_{12}(\varepsilon_{xx}\varepsilon_{yy} + \varepsilon_{xx}\varepsilon_{zz} + \varepsilon_{yy}\varepsilon_{zz}) + 4C_{44}(\varepsilon_{xy}^2 + \varepsilon_{xz}^2 + \varepsilon_{yz}^2)\right). \tag{4}$$

Note that all variables under the sign of integral in Eq. (4) are functions of $\mathbf{r}$. For ZB QDs embedded into a ZB matrix with lattice constant $a_{\text{matrix}}$, the tensor of local intrinsic strain is

$$\varepsilon_{ij}^{(0)}(\mathbf{r}) = \delta_{ij}\left(a(\mathbf{r}) - a_{\text{matrix}}\right)/a_{\text{matrix}}, \tag{5}$$

where $a(\mathbf{r})$ takes values of QD lattice constants inside QDs and is equal to $a_{\text{matrix}}$ outside QDs.

### B. Wurtzite quantum dots

Following standard notation, it is assumed in the following that the *z*-axis is the axis of sixfold rotational symmetry in WZ materials. In crystals with WZ symmetry, there are five linearly independent elastic constants: $\lambda_{xxxx} = C_{11}$, $\lambda_{zzzz} = C_{33}$, $\lambda_{xxyy} = C_{12}$, $\lambda_{xxzz} = C_{13}$, and $\lambda_{xzxz} = C_{44}$. Thus, the elastic energy (1) can be written as

$$F_{\text{elastic}} = \frac{1}{2}\int_V d\mathbf{r}\Big(C_{11}(\varepsilon_{xx}^2 + \varepsilon_{yy}^2) + C_{33}\varepsilon_{zz}^2 + 2C_{12}\varepsilon_{xx}\varepsilon_{yy} + 2C_{13}\varepsilon_{zz}(\varepsilon_{xx} + \varepsilon_{yy}) \\ + 4C_{44}(\varepsilon_{xz}^2 + \varepsilon_{yz}^2) + 2(C_{11} - C_{12})\varepsilon_{xy}^2\Big). \tag{6}$$

Note that all variables in the integrand of Eq. (6) are functions of $\mathbf{r}$. For WZ QDs embedded in a WZ matrix with lattice constants $a_{\text{matrix}}$ and $c_{\text{matrix}}$, the tensor of local intrinsic strain is

$$\varepsilon_{ij}^{(0)}(\mathbf{r}) = (\delta_{ij} - \delta_{iz}\delta_{jz})\left(a(\mathbf{r}) - a_{\text{matrix}}\right)/a_{\text{matrix}} + \delta_{iz}\delta_{jz}\left(c(\mathbf{r}) - c_{\text{matrix}}\right)/c_{\text{matrix}}, \tag{7}$$

where $a(\mathbf{r})$ and $c(\mathbf{r})$ take values of the QD lattice constants inside the QDs and are equal to $a_{\text{matrix}}$ and $c_{\text{matrix}}$, respectively, outside the QDs.

### III. PIEZOELECTRIC FIELD IN QUANTUM-DOT HETEROSTRUCTURES

Under an applied stress, some semiconductors develop an electric moment whose magnitude is proportional to the stress. The strain-induced polarization $\mathbf{P}^{\text{strain}}$ can be related to the strain tensor $\varepsilon_{lm}$ using the piezoelectric coefficients $e_{ilm}$ as follows,

$$P_i^{\text{strain}}(\mathbf{r}) = \sum_{lm} e_{ilm}(\mathbf{r}) \varepsilon_{lm}(\mathbf{r}),  \qquad (8)$$

where the indices *ilm* run over the spatial coordinates $x$, $y$, and $z$. Converting from tensor notation to matrix notation, Eq. (8) can be written as

$$P_i^{\text{strain}}(\mathbf{r}) = \sum_{k=1}^{6} e_{ik}(\mathbf{r}) \varepsilon_k(\mathbf{r}),  \qquad (9)$$

where $\{\varepsilon_{xx}, \varepsilon_{yy}, \varepsilon_{zz}, (\varepsilon_{yz}, \varepsilon_{zy}), (\varepsilon_{xz}, \varepsilon_{zx}), (\varepsilon_{xy}, \varepsilon_{yx})\} \equiv \{\varepsilon_1, \varepsilon_2, \varepsilon_3, \varepsilon_4, \varepsilon_5, \varepsilon_6\}$ and

$$e_{ilm} = \begin{cases} e_{ik}, & k = 1, 2, 3; \\ \tfrac{1}{2} e_{ik}, & k = 4, 5, 6. \end{cases} \qquad (10)$$

WZ nitrides also exhibit spontaneous polarization, $\mathbf{P}^{\text{spont}}$, with polarity specified by the terminating anion or cation at the surface. The total polarization,

$$\mathbf{P}(\mathbf{r}) = \mathbf{P}^{\text{strain}}(\mathbf{r}) + \mathbf{P}^{\text{spont}}(\mathbf{r}), \qquad (11)$$

leads to the appearance of an electrostatic piezoelectric potential, $V_p$. In the absence of external charges, the piezoelectric potential is found by solving the Maxwell equation:

$$\nabla \cdot \mathbf{D}(\mathbf{r}) = 0, \qquad (12)$$

where the displacement vector $\mathbf{D}$ in the system is

$$\mathbf{D}(\mathbf{r}) = -\hat{\varepsilon}_{\text{stat}}(\mathbf{r}) \nabla V_p(\mathbf{r}) + 4\pi \mathbf{P}(\mathbf{r}). \qquad (13)$$

In Eq. (13) $\hat{\varepsilon}_{\text{stat}}$ is the static dielectric tensor and $\mathbf{P}(\mathbf{r})$ is given by Eq. (11).

### A. Zinc-blende quantum dots

In crystals with ZB symmetry, only off-diagonal terms of the strain tensor give rise to the polarization. In component form,

$$\begin{aligned} P_x &= e_{14} \varepsilon_{yz}, \\ P_y &= e_{14} \varepsilon_{xz}, \\ P_z &= e_{14} \varepsilon_{xy}, \end{aligned} \qquad (14)$$

where $e_{14}$ is the only independent piezoelectric coefficient that survives, due to the ZB symmetry. The dielectric tensor in ZB materials reduces to a constant

$$\hat{\varepsilon}_{\text{stat}} = \begin{pmatrix} \varepsilon_{\text{stat}} & 0 & 0 \\ 0 & \varepsilon_{\text{stat}} & 0 \\ 0 & 0 & \varepsilon_{\text{stat}} \end{pmatrix}. \qquad (15)$$

### B. Wurtzite quantum dots

Self-assembled WZ QDs usually grow along the *z*-axis. In this case, only the *z*-component of the spontaneous polarization is nonzero: $P_z^{\text{spont}} \equiv P_{\text{sp}}$, where $P_{\text{sp}}$ is a specific constant for each material in a QD heterostructure. In crystals with WZ symmetry, the three distinct piezoelectric coefficients are $e_{15}$, $e_{31}$, and $e_{33}$. Thus, the polarization is given in component form by

$$P_x = e_{15}\varepsilon_{xz},$$
$$P_y = e_{15}\varepsilon_{yz}, \quad (16)$$
$$P_z = e_{31}(\varepsilon_{xx} + \varepsilon_{yy}) + e_{33}\varepsilon_{zz} + P_{sp}.$$

As seen from Eq. (16), both diagonal and off-diagonal terms of the strain tensor generate a built-in field in WZ QDs. The dielectric tensor in WZ materials has the following form

$$\hat{\varepsilon}_{stat} = \begin{pmatrix} \varepsilon_{stat}^{\perp} & 0 & 0 \\ 0 & \varepsilon_{stat}^{\perp} & 0 \\ 0 & 0 & \varepsilon_{stat}^{\parallel} \end{pmatrix}. \quad (17)$$

## IV. ELECTRON AND HOLE STATES IN STRAINED QUANTUM-DOT HETEROSTRUCTURES

Since both GaN and AlN have large band gaps (see Table I), we neglect coupling between the conduction and valence bands and consider separate one-band electron and six-band hole Hamiltonians. We also use proper operator ordering in the multi-band Hamiltonians, as is essential for an accurate description of QD heterostructures [19, 20].

TABLE I. Parameters of WZ GaN, WZ AlN, ZB GaN, and ZB AlN. Parameters, for which the source is not indicated explicitly, are taken from Ref. [13].

| Parameters | WZ GaN | WZ AlN | Parameters | ZB GaN | ZB AlN |
|---|---|---|---|---|---|
| $a$ (nm) | 3.189 | 3.112 | $a$ (nm) | 4.50 | 4.38 |
| $c$ (nm) | 5.185 | 4.982 | $E_g$ (eV) | 3.26 [7] | 4.9 |
| $E_g$ (eV) | 3.475 [2] | 6.23 | $\Delta_{so}$ (eV) | 0.017 | 0.019 |
| $\Delta_{cr}$ (eV) | 0.019 | $-0.164$ | $m_e$ | 0.15 | 0.25 |
| $\Delta_{so}$ (eV) | 0.014 | 0.019 | $\gamma_1$ | 2.67 | 1.92 |
| $m_e^{\parallel}$ | 0.20 | 0.28 | $\gamma_2$ | 0.75 | 0.47 |
| $m_e^{\perp}$ | 0.20 | 0.32 | $\gamma_3$ | 1.10 | 0.85 |
| $A_1$ | $-6.56$ | $-3.95$ | $E_P$ (eV) | 25.0 | 27.1 |
| $A_2$ | $-0.91$ | $-0.27$ | VBO (eV) | 0 | $-0.8$ |
| $A_3$ | 5.65 | 3.68 | $a_c$ (eV) | $-2.2$ | $-6.0$ |
| $A_4$ | $-2.83$ | $-1.84$ | $a_v$ (eV) | $-5.2$ | $-3.4$ |
| $A_5$ | $-3.13$ | $-1.95$ | $b$ (eV) | $-2.2$ | $-1.9$ |
| $A_6$ | $-4.86$ | $-2.91$ | $d$ (eV) | $-3.4$ | $-10$ |
| $E_P$ (eV) | 14.0 | 14.5 | $C_{11}$ (GPa) | 293 | 304 |
| VBO (eV) | 0 | $-0.8$ | $C_{12}$ (GPa) | 159 | 160 |
| $a_c^{\parallel}$ (eV) | $-9.5$ | $-12.0$ | $C_{44}$ (GPa) | 155 | 193 |
| $a_c^{\perp}$ (eV) | $-8.2$ | $-5.4$ | $e_{14}$ (C/m$^2$) | 0.50 [14] | 0.59 [14] |
| $D_1$ (eV) | $-3.0$ | $-3.0$ | $\varepsilon_{stat}$ | 9.7 [15] | 9.7 [15] |
| $D_2$ (eV) | 3.6 | 3.6 | $\varepsilon_{opt}$ | 5.3 [15] | 5.3 [15] |
| $D_3$ (eV) | 8.82 | 9.6 | $n$ | 2.29 [16] | |

| | | |
|---|---|---|
| $D_4$ (eV) | −4.41 | −4.8 |
| $D_5$ (eV) | −4.0 | −4.0 |
| $D_6$ (eV) | −5.1 | −5.1 |
| $C_{11}$ (GPa) | 390 | 396 |
| $C_{12}$ (GPa) | 145 | 137 |
| $C_{13}$ (GPa) | 106 | 108 |
| $C_{33}$ (GPa) | 398 | 373 |
| $C_{44}$ (GPa) | 105 | 116 |
| $e_{15}$ (C/m$^2$) | −0.49 [17] | −0.60 [17] |
| $e_{31}$ (C/m$^2$) | −0.49 [17] | −0.60 [17] |
| $e_{33}$ (C/m$^2$) | 0.73 [17] | 1.46 [17] |
| $P_{sp}$ (C/m$^2$) | −0.029 [17] | −0.081 [17] |
| $\varepsilon_{stat}^{\parallel}$ | 10.01 [18] | 8.57 [18] |
| $\varepsilon_{stat}^{\perp}$ | 9.28 [18] | 8.67 [18] |
| $\varepsilon_{opt}$ | 5.29 [18] | 4.68 [18] |
| $n$ | 2.29 [16] | |

Electron states are eigenstates of the one-band envelope-function equation:

$$\hat{H}_e \Psi_e = E_e \Psi_e, \qquad (18)$$

where $\hat{H}_e$, $\Psi_e$, and $E_e$ are the electron Hamiltonian, the envelope wave function and the energy, respectively. Each electron energy level is twofold degenerate with respect to spin. The two microscopic electron wave functions corresponding to an eigenenergy $E_e$ are

$$\begin{bmatrix} \psi_e = \Psi_e |S\rangle |\uparrow\rangle; \\ \psi_e = \Psi_e |S\rangle |\downarrow\rangle, \end{bmatrix} \qquad (19)$$

where $|S\rangle$ is Bloch function of the conduction band and $|\uparrow\rangle$, $|\downarrow\rangle$ are electron spin functions. The electron Hamiltonian $\hat{H}_e$ can be written as

$$\hat{H}_e = \hat{H}_S(\mathbf{r}_e) + H_e^{(\varepsilon)}(\mathbf{r}_e) + E_c(\mathbf{r}_e) + eV_p(\mathbf{r}_e), \qquad (20)$$

where $\hat{H}_S$ is the kinetic part of the microscopic Hamiltonian unit-cell averaged by the Bloch function $|S\rangle$, $H_e^{(\varepsilon)}$ is the strain-dependent part of the electron Hamiltonian, $E_c$ is the energy of unstrained conduction band edge, $e$ is the absolute value of electron charge, and $V_p$ is the piezoelectric potential.

Hole states are eigenstates of the six-band envelope-function equation:

$$\hat{H}_h \Psi_h = E_h \Psi_h, \qquad (21)$$

where $\hat{H}_h$ is 6×6 matrix of the hole Hamiltonian, $\Psi_h$ is 6-component column of the hole envelope wave function, and $E_h$ is the hole energy. The microscopic hole wave function corresponding to an eigenenergy $E_h$ is

$$\psi_h = \left( |X\rangle|\uparrow\rangle,\ |Y\rangle|\uparrow\rangle,\ |Z\rangle|\uparrow\rangle,\ |X\rangle|\downarrow\rangle,\ |Y\rangle|\downarrow\rangle,\ |Z\rangle|\downarrow\rangle \right) \cdot \Psi_h, \quad (22)$$

where $|X\rangle$, $|Y\rangle$, and $|Z\rangle$ are Bloch function of the valence band and $|\uparrow\rangle$, $|\downarrow\rangle$ are spin functions of the missing electron. The hole Hamiltonian $\hat{H}_h$ can be written as

$$\hat{H}_h = \begin{pmatrix} \hat{H}_{XYZ}(\mathbf{r}_h) + H_h^{(\varepsilon)}(\mathbf{r}_h) & 0 \\ 0 & \hat{H}_{XYZ}(\mathbf{r}_h) + H_h^{(\varepsilon)}(\mathbf{r}_h) \end{pmatrix} + E_v(\mathbf{r}_h) + eV_p(\mathbf{r}_h) + H_{so}(\mathbf{r}_h). \quad (23)$$

$\hat{H}_{XYZ}$ is a $3\times 3$ matrix of the kinetic part of the microscopic Hamiltonian, unit-cell averaged by the Bloch functions $|X\rangle$, $|Y\rangle$, and $|Z\rangle$ (the crystal-field splitting is also included in $\hat{H}_{XYZ}$ for WZ QDs). $H_h^{(\varepsilon)}$ is a $3\times 3$ matrix of the strain-dependent part of the hole Hamiltonian, $E_v$ is the energy of the unstrained valence band edge, $e$ is the absolute value of the electron charge and $V_p$ is the piezoelectric potential. The last term in Eq. (23) is the Hamiltonian of spin-orbit interaction [19]:

$$H_{so}(\mathbf{r}) = \frac{\Delta_{so}(\mathbf{r})}{3} \begin{pmatrix} -1 & -i & 0 & 0 & 0 & 1 \\ i & -1 & 0 & 0 & 0 & -i \\ 0 & 0 & -1 & -1 & i & 0 \\ 0 & 0 & -1 & -1 & i & 0 \\ 0 & 0 & -i & -i & -1 & 0 \\ 1 & i & 0 & 0 & 0 & -1 \end{pmatrix}, \quad (24)$$

where $\Delta_{so}$ is the spin-orbit splitting energy.

### A.  Zinc-blende quantum dots

For ZB QDs, the first term in the electron Hamiltonian (20) has the form

$$\hat{H}_S(\mathbf{r}) = \frac{\hbar^2}{2m_0} \hat{\mathbf{k}} \frac{1}{m_e(\mathbf{r})} \hat{\mathbf{k}}, \quad (25)$$

where $\hbar$ is Planck's constant, $m_0$ is the free-electron mass, $\hat{\mathbf{k}} = -i\nabla$ is the wave vector operator and $m_e$ is the electron effective mass in units of $m_0$. The strain-dependent part of the electron Hamiltonian (20) is

$$H_e^{(\varepsilon)}(\mathbf{r}) = a_c(\mathbf{r})\left(\varepsilon_{xx}(\mathbf{r}) + \varepsilon_{yy}(\mathbf{r}) + \varepsilon_{zz}(\mathbf{r})\right), \quad (26)$$

where $a_c$ is the conduction-band deformation potential and $\varepsilon_{ij}$ is the strain tensor.

The matrix $\hat{H}_{XYZ}$ entering the hole Hamiltonian (23) is given by [19]

$$\hat{H}_{XYZ} = -\frac{\hbar^2}{2m_0} \begin{pmatrix} \hat{k}_x \beta_l \hat{k}_x + \hat{\mathbf{k}}_x^\perp \beta_h \hat{\mathbf{k}}_x^\perp & 3\left(\hat{k}_x \gamma_3^+ \hat{k}_y + \hat{k}_y \gamma_3^- \hat{k}_x\right) & 3\left(\hat{k}_x \gamma_3^+ \hat{k}_z + \hat{k}_z \gamma_3^- \hat{k}_x\right) \\ 3\left(\hat{k}_x \gamma_3^- \hat{k}_y + \hat{k}_y \gamma_3^+ \hat{k}_x\right) & \hat{k}_y \beta_l \hat{k}_y + \hat{\mathbf{k}}_y^\perp \beta_h \hat{\mathbf{k}}_y^\perp & 3\left(\hat{k}_y \gamma_3^+ \hat{k}_z + \hat{k}_z \gamma_3^- \hat{k}_y\right) \\ 3\left(\hat{k}_x \gamma_3^- \hat{k}_z + \hat{k}_z \gamma_3^+ \hat{k}_x\right) & 3\left(\hat{k}_y \gamma_3^- \hat{k}_z + \hat{k}_z \gamma_3^+ \hat{k}_y\right) & \hat{k}_z \beta_l \hat{k}_z + \hat{\mathbf{k}}_z^\perp \beta_h \hat{\mathbf{k}}_z^\perp \end{pmatrix}, \quad (27)$$

where $\hat{\mathbf{k}}_i^\perp = \hat{\mathbf{k}} - \hat{\mathbf{k}}_i$ ($i = x, y, z$),

$$\begin{aligned} \beta_l &= \gamma_1 + 4\gamma_2, \\ \beta_h &= \gamma_1 - 2\gamma_2, \\ \gamma_3^+ &= (2\gamma_2 + 6\gamma_3 - \gamma_1 - 1)/3, \\ \gamma_3^- &= (-2\gamma_2 + \gamma_1 + 1)/3. \end{aligned} \quad (28)$$

In Eq. (28), $\gamma_1$, $\gamma_2$, and $\gamma_3$ are the Luttinger-Kohn parameters of the valence band. The strain-dependent part, $H_h^{(\varepsilon)}$, of the hole Hamiltonian (23) can be written as [21]

$$H_h^{(\varepsilon)} = -a_v\left(\varepsilon_{xx} + \varepsilon_{yy} + \varepsilon_{zz}\right) + \begin{pmatrix} b(2\varepsilon_{xx} - \varepsilon_{yy} - \varepsilon_{zz}) & \sqrt{3}\,d\,\varepsilon_{xy} & \sqrt{3}\,d\,\varepsilon_{xz} \\ \sqrt{3}\,d\,\varepsilon_{xy} & b(2\varepsilon_{yy} - \varepsilon_{xx} - \varepsilon_{zz}) & \sqrt{3}\,d\,\varepsilon_{yz} \\ \sqrt{3}\,d\,\varepsilon_{xz} & \sqrt{3}\,d\,\varepsilon_{yz} & b(2\varepsilon_{zz} - \varepsilon_{xx} - \varepsilon_{yy}) \end{pmatrix}, \quad (29)$$

where $a_v$, $b$, and $d$ are the hydrostatic and two shear valence-band deformation potentials, respectively. Note that all parameters in Eqs. (27) and (29) are coordinate-dependent for QD heterostructures.

**B.    Wurtzite quantum dots**

For WZ QDs, the first term in the electron Hamiltonian (20) has the form

$$\hat{H}_S(\mathbf{r}) = \frac{\hbar^2}{2m_0}\left(\hat{k}_z \frac{1}{m_e^\parallel(\mathbf{r})} \hat{k}_z + \hat{\mathbf{k}}_z^\perp \frac{1}{m_e^\perp(\mathbf{r})} \hat{\mathbf{k}}_z^\perp\right), \quad (30)$$

where $m_e^\parallel$ and $m_e^\perp$ are electron effective masses in units of $m_0$ and $\hat{\mathbf{k}}_z^\perp = \hat{\mathbf{k}} - \hat{\mathbf{k}}_z$. The strain-dependent part of the electron Hamiltonian (20) is

$$H_e^{(\varepsilon)}(\mathbf{r}) = a_c^\parallel(\mathbf{r})\,\varepsilon_{zz}(\mathbf{r}) + a_c^\perp(\mathbf{r})\left(\varepsilon_{xx}(\mathbf{r}) + \varepsilon_{yy}(\mathbf{r})\right), \quad (31)$$

where $a_c^\parallel$ and $a_c^\perp$ are conduction-band deformation potentials.

The matrix $\hat{H}_{XYZ}$ entering the hole Hamiltonian (23) is given by [20]

$$\hat{H}_{XYZ} = \frac{\hbar^2}{2m_0}\begin{pmatrix} \hat{k}_x L_1 \hat{k}_x + \hat{k}_y M_1 \hat{k}_y + \hat{k}_z M_2 \hat{k}_z & \hat{k}_x N_1 \hat{k}_y + \hat{k}_y N_1' \hat{k}_x & \hat{k}_x N_2 \hat{k}_z + \hat{k}_z N_2' \hat{k}_x \\ \hat{k}_y N_1 \hat{k}_x + \hat{k}_x N_1' \hat{k}_y & \hat{k}_x M_1 \hat{k}_x + \hat{k}_y L_1 \hat{k}_y + \hat{k}_z M_2 \hat{k}_z & \hat{k}_y N_2 \hat{k}_z + \hat{k}_z N_2' \hat{k}_y \\ \hat{k}_z N_2 \hat{k}_x + \hat{k}_x N_2' \hat{k}_z & \hat{k}_z N_2 \hat{k}_y + \hat{k}_y N_2' \hat{k}_z & \hat{k}_x M_3 \hat{k}_x + \hat{k}_y M_3 \hat{k}_y + \hat{k}_z L_2 \hat{k}_z - \delta_{cr} \end{pmatrix}, \quad (32)$$

where

$$\begin{aligned} L_1 &= A_2 + A_4 + A_5, \quad L_2 = A_1, \\ M_1 &= A_2 + A_4 - A_5, \quad M_2 = A_1 + A_3, \quad M_3 = A_2, \\ N_1 &= 3A_5 - (A_2 + A_4) + 1, \quad N_1' = -A_5 + A_2 + A_4 - 1, \\ N_2 &= 1 - (A_1 + A_3) + \sqrt{2} A_6, \quad N_2' = A_1 + A_3 - 1, \\ \delta_{cr} &= 2m_0 \Delta_{cr}/\hbar^2. \end{aligned} \quad (33)$$

In Eq. (33), $A_k$ ($k = 1, \ldots, 6$) are Rashba-Sheka-Pikus parameters of the valence band and $\Delta_{cr}$ is the crystal-field splitting energy. The strain-dependent part $H_h^{(\varepsilon)}$ of the hole Hamiltonian (23) can be written as [20]

$$H_h^{(\varepsilon)} = \begin{pmatrix} l_1\varepsilon_{xx} + m_1\varepsilon_{yy} + m_2\varepsilon_{zz} & n_1\varepsilon_{xy} & n_2\varepsilon_{xz} \\ n_1\varepsilon_{xy} & m_1\varepsilon_{xx} + l_1\varepsilon_{yy} + m_2\varepsilon_{zz} & n_2\varepsilon_{yz} \\ n_2\varepsilon_{xz} & n_2\varepsilon_{yz} & m_3(\varepsilon_{xx} + \varepsilon_{yy}) + l_2\varepsilon_{zz} \end{pmatrix}, \quad (34)$$

where

$$\begin{aligned} l_1 &= D_2 + D_4 + D_5, \quad l_2 = D_1, \\ m_1 &= D_2 + D_4 - D_5, \quad m_2 = D_1 + D_3, \quad m_3 = D_2, \\ n_1 &= 2D_5, \quad n_2 = \sqrt{2}D_6. \end{aligned} \quad (35)$$

In Eq. (35), $D_k$ ($k = 1, \ldots, 6$) are valence-band deformation potentials. Note, that all parameters in Eqs. (32) and (34) are coordinate-dependent for QD heterostructures.

## V. COULOMB POTENTIAL ENERGY IN QUANTUM-DOT HETEROSTRUCTURES

The Coulomb potential energy of the electron-hole system in a QD heterostructure is [22]

$$U(\mathbf{r}_e, \mathbf{r}_h) = U_{int}(\mathbf{r}_e, \mathbf{r}_h) + U_{s\text{-}a}(\mathbf{r}_e) + U_{s\text{-}a}(\mathbf{r}_h). \quad (36)$$

In Eq. (36), $U_{int}(\mathbf{r}_e, \mathbf{r}_h)$ is the electron-hole interaction energy, which is the solution of the Poisson equation:

$$\nabla_{\mathbf{r}_h}\left(\varepsilon_{opt}(\mathbf{r}_h)\nabla_{\mathbf{r}_h}U_{int}(\mathbf{r}_e, \mathbf{r}_h)\right) = \frac{e^2}{\varepsilon_0}\delta(\mathbf{r}_e - \mathbf{r}_h), \quad (37)$$

where $\varepsilon_{opt}$ is the optical dielectric constant, $\varepsilon_0$ is the permittivity of free space, and $\delta$ is the Dirac delta function. The second and third terms on the right-hand side of Eq. (36) are the electron and hole self-interaction energies, defined as

$$U_{s\text{-}a}(\mathbf{r}) = -\frac{1}{2}\lim_{\mathbf{r}' \to \mathbf{r}}\left[U_{int}(\mathbf{r}, \mathbf{r}') - U_{int}^{bulk}(\mathbf{r}, \mathbf{r}')\right], \quad (38)$$

where $U_{int}^{bulk}(\mathbf{r}, \mathbf{r}')$ is the local bulk solution of Eq. (37), i.e.

$$U_{int}^{bulk}(\mathbf{r}, \mathbf{r}') = -\frac{e^2}{4\pi\varepsilon_0\varepsilon_{opt}(\mathbf{r})|\mathbf{r} - \mathbf{r}'|}. \quad (39)$$

It should be pointed out that an infinite discontinuity in the self-interaction energy (38) arises at the boundaries between different materials of the heterostructure, when the optical dielectric constant $\varepsilon_{opt}(\mathbf{r})$ changes abruptly form its value in one material to its value in the adjacent material. This theoretical difficulty can be overcome easily by considering a transitional layer between the two materials, where $\varepsilon_{opt}(\mathbf{r})$ changes gradually between its values in different materials. The thickness of the transitional layer in self-assembled QDs depends on the growth parameters and is usually of order of one monolayer.

## VI. EXCITON STATES, OSCILLATOR STRENGTHS AND RADIATIVE DECAY TIMES

In the strong confinement regime, the exciton wave function $\psi_{exc}$ can be approximated by the wave function of the electron-hole pair:

$$\psi_{exc}(\mathbf{r}_e, \mathbf{r}_h) = \psi_e^*(\mathbf{r}_e) \psi_h(\mathbf{r}_h), \qquad (40)$$

and the exciton energy $E_{exc}$ can be calculated considering the Coulomb potential energy (36) as a perturbation:

$$E_{exc} = E_e - E_h + \int_V d\mathbf{r}_e \int_V d\mathbf{r}_h U(\mathbf{r}_e, \mathbf{r}_h) |\psi_{exc}(\mathbf{r}_e, \mathbf{r}_h)|^2. \qquad (41)$$

The electron and hole wave functions $\psi_e$ and $\psi_h$ in Eq. (40) are given by Eqs. (19) and (22), correspondingly. In Eq. (41), $E_e$ and $E_h$ are electron and hole energies, and $V$ is the total volume of the system.

The oscillator strength $f$ of the exciton [Eqs. (40), (41)] can be calculated as

$$f = \frac{2\hbar^2}{m_0 E_{exc}} \sum_\alpha \left| \int_V d\mathbf{r}\, \psi_e^*(\mathbf{r}) (\mathbf{e}, \hat{\mathbf{k}}) \psi_h^{(\alpha)}(\mathbf{r}) \right|^2, \qquad (42)$$

where $\mathbf{e}$ is the polarization of incident light, $\hat{\mathbf{k}} = -i\nabla$ is the wave vector operator, and $\alpha$ denotes different hole wave functions corresponding to the same degenerate hole energy level $E_h$. To calculate the oscillator strength $f$, the integral over the volume $V$ in Eq. (42) should be represented as a sum of integrals over unit cells contained in the volume $V$. When integrating over the volume of each unit cell, envelope wave functions $\Psi_e$ and $\Psi_h$ are treated as specific for each unit cell constants. In this case, each integral over the volume of a unit cell is proportional to the constant:

$$\langle S | \hat{k}_i | I \rangle = \delta_{i,I} \sqrt{\frac{m_0 E_P}{2\hbar^2}}, \qquad (43)$$

which is equal for each unit cell of the same material. In Eq. (43) $i, I = X, Y, Z$; $\delta_{i,I}$ is the Kronecker delta symbol; and $E_P$ is the Kane energy.

The oscillator strength $f$ not only defines the strength of absorption lines, but also relates to the radiative decay time $\tau$ [23]:

$$\tau = \frac{2\pi\varepsilon_0 m_0 c^3 \hbar^2}{n e^2 E_{exc}^2 f}, \qquad (44)$$

where $\varepsilon_0$, $m_0$, $c$, $\hbar$, and $e$ are fundamental physical constants with their usual meaning and $n$ is the refractive index.

## VII. RESULTS OF THE CALCULATION FOR DIFFERENT GAN/ALN QUANTUM DOTS

The theory described in Sections II–VI is applied in this section to describe excitonic properties of strained WZ and ZB GaN/AlN and WZ GaN/Al$_{0.15}$Ga$_{0.85}$N QDs. We consider the following three kinds of single GaN QDs with variable QD height $H$: (i) WZ GaN/AlN QDs [see Fig. 1(a)] with the thickness of the wetting layer $w = 0.5$ nm, QD bottom diameter $D_B = 5(H - w)$, and QD top diameter $D_T = H - w$ [1, 2]; (ii) ZB GaN/AlN QDs [see Fig. 1(b)] with $w = 0.5$ nm, QD bottom base length $D_B = 10(H - w)$, and QD top base length $D_T = 8.6(H - w)$ [7, 8]; (iii) Disk-shaped WZ GaN/Al$_{0.15}$Ga$_{0.85}$N QDs with $w = 0$ and QD diameter $D = 3H$ [3, 4]. Material parameters used in our calculations are listed in Table I. A linear interpolation is used to find the material parameters of WZ Al$_{0.15}$Ga$_{0.85}$N from the material parameters of WZ GaN and WZ AlN.

It should be pointed out that WZ GaN/AlN and ZB GaN/AlN QDs are grown as 3-D arrays of GaN QDs in the AlN matrix [1, 2, 5, 6], while WZ GaN/Al$_x$Ga$_{1-x}$N QDs are grown as uncapped 2-D arrays of GaN QDs on the Al$_x$Ga$_{1-x}$N layer [3, 4]. While the distance between GaN QDs in a plane perpendicular to the growth direction is sufficiently large and should not influence optical properties of the system, the distance between GaN QDs along the growth direction can be made rather small. In the latter case, a vertical correlation is observed between GaN QDs, which can also affect optical properties of the system. The theory described in Sections II–VI can be directly applied to describe vertically correlated WZ GaN/AlN and ZB GaN/AlN QDs. Since we are mainly interested in the properties of excitons in the ground and lowest excited states, here, we consider single GaN QDs in the AlN matrix. Within our model, uncapped GaN QDs on the Al$_x$Ga$_{1-x}$N layer can be considered as easily as GaN QDs in the Al$_x$Ga$_{1-x}$N matrix. In the following we consider GaN QDs in the Al$_x$Ga$_{1-x}$N matrix to facilitate comparison with WZ GaN/AlN and ZB GaN/AlN QDs.

The strain tensor in WZ and ZB GaN/AlN and WZ GaN/Al$_{0.15}$Ga$_{0.85}$N QDs has been calculated by minimizing the elastic energy given by Eq. (4) for WZ QDs and the one given by Eq. (6) for ZB QDs with respect to the displacement vector $\mathbf{u}(\mathbf{r})$. We have carried out the numerical minimization of the elastic energy $F_{\text{elastic}}$ by, first, employing the finite-element method to evaluate the integrals $F_{\text{elastic}}$ as a function of $u_i(\mathbf{r_n})$, where $i = x, y, z$ and $\mathbf{n}$ numbers our finite-elements; second, transforming our extremum problem to a system of linear equations $\partial F_{\text{elastic}} / \partial u_i(\mathbf{r_n}) = 0$; and third, solving the obtained system of linear equations with the boundary conditions that $\mathbf{u}(\mathbf{r})$ vanishes sufficiently far from the QD.

Using the calculated strain tensor, we compute the piezoelectric potential for WZ and ZB GaN/AlN and WZ GaN/Al$_{0.15}$Ga$_{0.85}$N QDs by solving the Maxwell equation [Eqs. (12), (13)] with the help of the finite-difference method. Figs. 2(a) and 2(b) show the piezoelectric potential in WZ and ZB GaN/AlN QDs with height 3 nm, correspondingly. It is seen that the magnitude of the piezoelectric potential in a WZ GaN/AlN QD is about 10 times its magnitude in a ZB GaN/AlN QD. Moreover, the piezoelectric potential in the WZ QD has maxima near the QD top and bottom, while the maxima of the piezoelectric potential in the ZB QD lie outside the QD. The above facts explain why the piezoelectric field has a strong effect on the excitonic properties of WZ GaN/AlN QDs, while it has very little effect on those in ZB GaN/AlN QDs.

Both strain and piezoelectric fields modify bulk conduction and valence band edges of GaN QDs [see Eqs. (20) and (23)]. As seen from Figs. 3(a) and (b), the piezoelectric potential in a WZ GaN/AlN QD tilts conduction and valence band edges along the z-axis in such a way that it becomes energetically favorable for the electron to be located near the QD top and for the hole to be located in the wetting layer, near to the QD bottom. On the other hand, it is seen from Figs. 4(a) and (b) that the deformation potential in a ZB GaN/AlN QD bends the valence band edge in the xy-plane in such a way that it creates a parabolic-like potential well that expells the hole from the QD side edges. Figs. 3 and 4 also show that the strain field pulls conduction and valence bands apart and significantly splits the valence band edge.

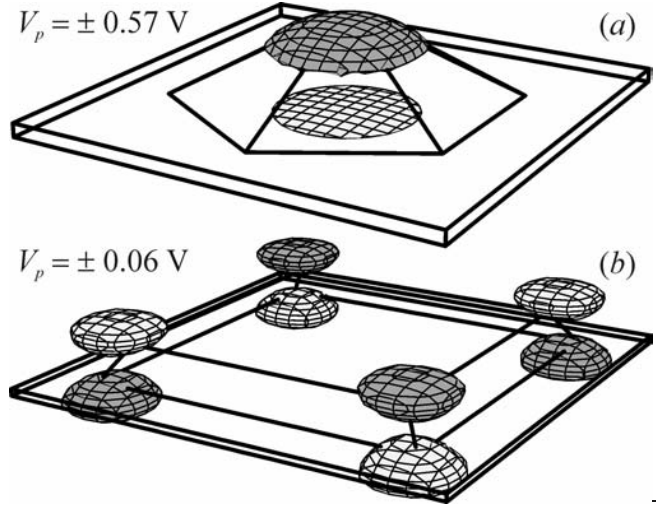

FIG. 2. Piezoelectric potential in WZ GaN/AlN (a) and ZB GaN/AlN (b) QDs with height 3 nm. The light and dark surfaces represent positive and negative values of the piezoelectric potential, correspondingly.

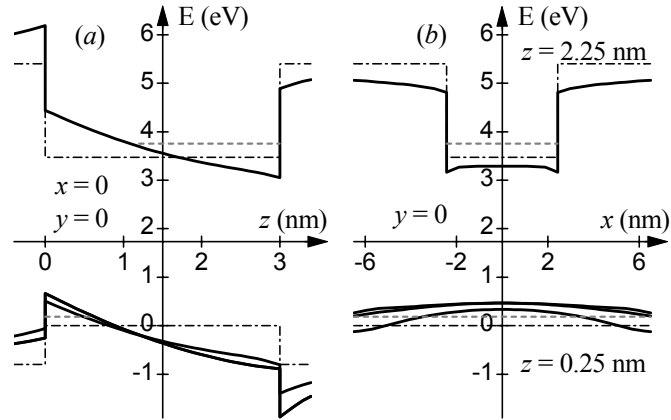

FIG. 3. Conduction and valence band edges along $z$-axis (a) and along $x$-axis (b) for WZ GaN/AlN QD with height 3 nm (solid lines). The valence band edge is split due to the strain and crystal fields. Dash-dotted lines show the conduction and valence band edges in the absence of strain and piezoelectric fields. Gray dashed lines show positions of electron and hole ground state energies.

Using the strain tensor and piezoelectric potential, electron and hole states have been calculated following Section IV. We have used the finite-difference method similar to that of Ref. [22] to find the lowest eigenstates of the electron envelope-function equation (18) and the hole envelope-function equation (21). The spin-orbit splitting energy in GaN and AlN is very small (see Table I); therefore, we follow the usual practice of neglecting it in the calculation of hole states in GaN QDs [9]. Fig. 5 presents four lowest electron states in WZ and ZB GaN/AlN

QDs with height 3 nm. Recalling the conduction band edge profiles (see Figs. 3 and 4), it becomes clear why the electron in the WZ GaN/AlN QD is pushed to the QD top, while the electron in the ZB GaN/AlN QD is distributed over the entire QD. The behavior of the four lowest hole states in WZ and ZB GaN/AlN QDs with height 3 nm (see Fig. 6) can be also predicted by looking at the valence band edge profiles shown in Figs. 3 and 4. Namely, the hole in the WZ GaN/AlN QD is pushed into the wetting layer and is located near the QD bottom, while the hole in the ZB GaN/AlN QD is expelled from the QD side edges. Due to the symmetry of QDs considered in this paper, the hole ground state energy is twofold degenerate, when the degeneracy by spin is not taken into account.

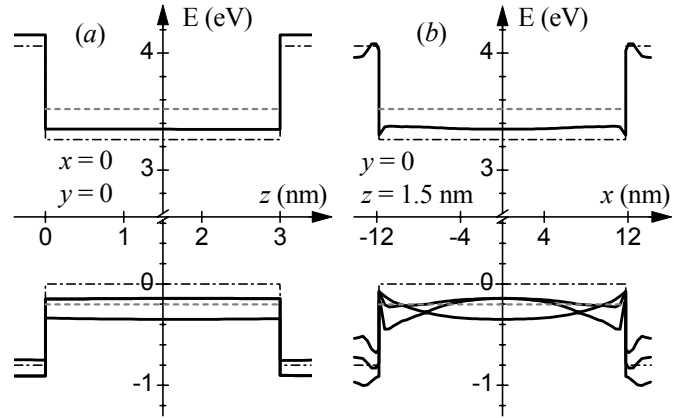

FIG. 4. Conduction and valence band edges along $z$-axis (a) and along $x$-axis (b) for ZB GaN/AlN QD with height 3 nm (solid lines). The valence band edge is split due to the strain field. Dash-dotted lines show the conduction and valence band edges in the absence of strain field. Gray dashed lines show positions of electron and hole ground state energies.

Both piezoelectric and strain fields are about seven times weaker in the WZ GaN/Al$_{0.15}$Ga$_{0.85}$N QD than they are in the WZ GaN/AlN QD. Therefore, conduction and valence band edges in WZ GaN/Al$_{0.15}$Ga$_{0.85}$N QDs do not differ significantly from their bulk positions and the electron and hole states are governed mainly by quantum confinement.

In the following we consider excitonic properties of WZ GaN/AlN, ZB GaN/AlN, and WZ GaN/Al$_{0.15}$Ga$_{0.85}$N QDs as a function of QD height. Fig. 7 shows electron and hole ground state energy levels in the three QDs. It is seen that the difference between the electron and hole energy levels decreases rapidly with increasing the QD height for WZ GaN/AlN QDs, unlike in two other kinds of QDs where the decrease is slower. The rapid decreasing of the electron-hole energy difference for WZ GaN/AlN QDs is explained by the fact that the magnitude of the piezoelectric potential increases linearly with increasing the QD height.

The exciton energy has been calculated using Eq. (41), where the Coulomb potential energy (36) has been computed with the help of a finite-difference method. Fig. 8 shows exciton ground state energy levels as a function of QD height for the three kinds of GaN QDs. Filled triangles, empty triangles, and filled circles show experimental points from Refs. [2], [8], and [4], correspondingly. The figure shows fair agreement between calculated exciton ground state energies and experimental data. It is seen that for WZ GaN/AlN QDs higher than 3nm, the exciton ground state energy drops below the bulk WZ GaN energy gap. Such a huge red-shift of the exciton ground state energy with respect to the bulk WZ GaN energy gap is attributed to the strong piezoelectric field in WZ GaN/AlN QDs. Due to the lower strength of the piezoelectric field in WZ GaN/Al$_{0.15}$Ga$_{0.85}$N QDs, the exciton ground state energy in these QDs becomes equal to the bulk WZ GaN energy gap only for a QD with height 4.5 nm. The piezoelectric field in ZB GaN/AlN QDs cannot significantly modify conduction and valence band edges, therefore the behavior of the exciton ground state energy with increasing QD height is mainly determined by the deformation potential and confinement.

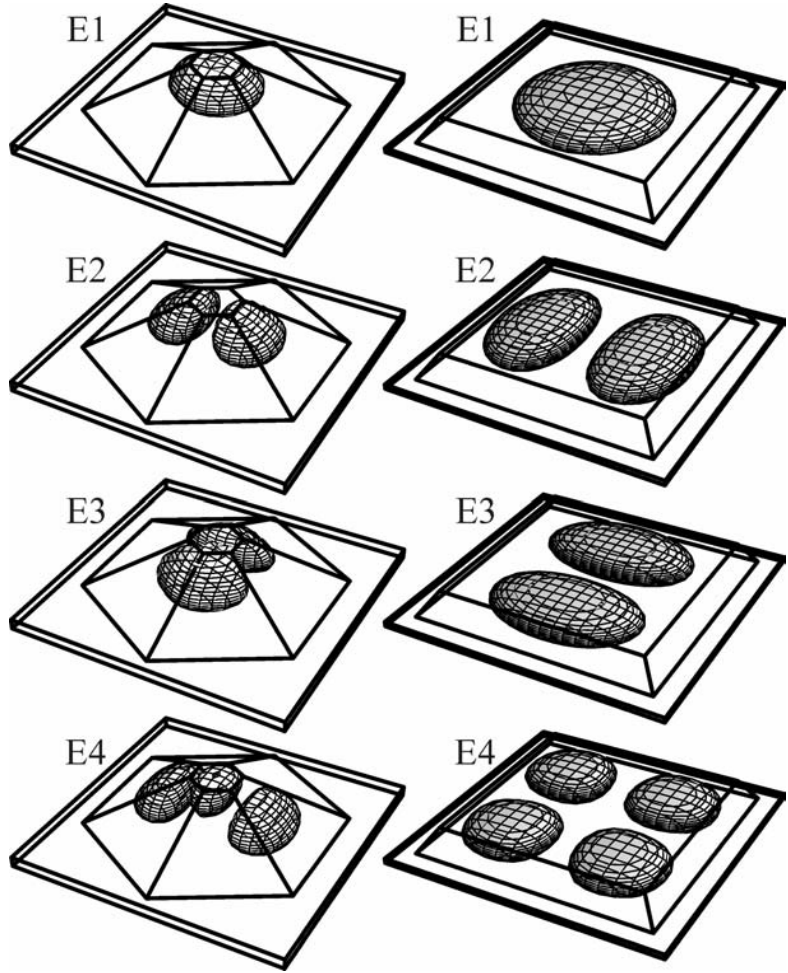

FIG. 5. Isosurfaces of probability density $|\psi_e|^2 = \rho$ for the four lowest electron states in WZ GaN/AlN (left panel) and ZB GaN/AlN (right panel) QDs with height 3 nm. $\rho$ is defined form the equation $\int_V |\psi_e(\mathbf{r})|^2 \theta\left[|\psi_e(\mathbf{r})|^2 - \rho\right] d\mathbf{r} = 0.8$, where $\theta$ is the Heaviside theta function. Energies of the electron states in the WZ QD are $E_1 = 3.752$ eV, $E_2 = 3.921$ eV, $E_3 = 3.962$ eV, and $E_4 = 4.074$ eV. Energies of the electron states in the ZB QD are $E_1 = 3.523$ eV, $E_2 = E_3 = 3.540$ eV, and $E_4 = 3.556$ eV.

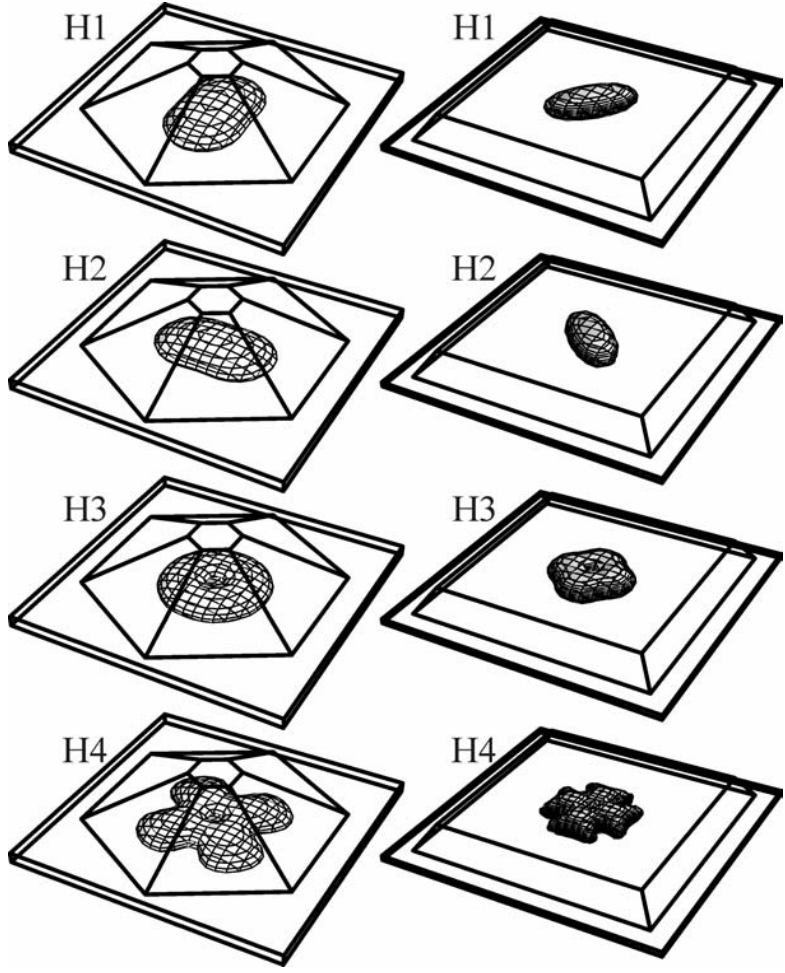

FIG. 6. Isosurfaces of probability density $|\psi_h|^2 = \rho$ for the four lowest hole states in WZ GaN/AlN (left panel) and ZB GaN/AlN (right panel) QDs with height 3 nm. $\rho$ is defined form the equation $\int_V |\psi_h(\mathbf{r})|^2 \theta\left[|\psi_h(\mathbf{r})|^2 - \rho\right] d\mathbf{r} = 0.8$, where $\theta$ is the Heaviside theta function. Energies of the hole states in the WZ QD are $E_1 = E_2 = 0.185$ eV, $E_3 = 0.171$ eV, and $E_4 = 0.156$ eV. Energies of the hole states in the ZB QD are $E_1 = E_2 = -0.202$ eV, $E_3 = -0.203$ eV, and $E_4 = -0.211$ eV.

Figs. 5 and 6 show that the electron and hole are spatially separated in WZ GaN/AlN QDs. This fact leads to very small oscillator strength (42) in those QDs. On the other hand, the charges are not separated in ZB GaN/AlN QDs, resulting in a large oscillator strength. An important physical quantity, the radiative decay time (44) is inversely proportional to the oscillator strength. Calculated radiative decay times of excitonic ground state transitions in the three kinds of GaN QDs are plotted in Fig. 9 as a function of QD height. The amplitude of the piezoelectric potential in WZ GaN/AlN and GaN/Al$_{0.15}$Ga$_{0.85}$N QDs increases with increasing the QD height. Therefore, the electron-hole separation also increases, the oscillator strength decreases, and the radiative decay time increases. The figure shows that the radiative decay time of the red-shifted transitions in WZ GaN/AlN QDs ($H > 3$ nm) is large and increases almost exponentially from 6.6 ns for QDs with height 3 nm to 1100 ns for QDs with height 4.5 nm. In WZ GaN/Al$_{0.15}$Ga$_{0.85}$N QDs, the radiative decay time and its increase with QD height are much smaller than those in WZ GaN/AlN QDs. The radiative decay time in ZB GaN/AlN QDs is found to be of order 0.3 ns and almost independent of QD height. Filled and empty triangles in Fig. 9 represent experimental points of Ref. [24], which appear to be in good agreement with our calculations.

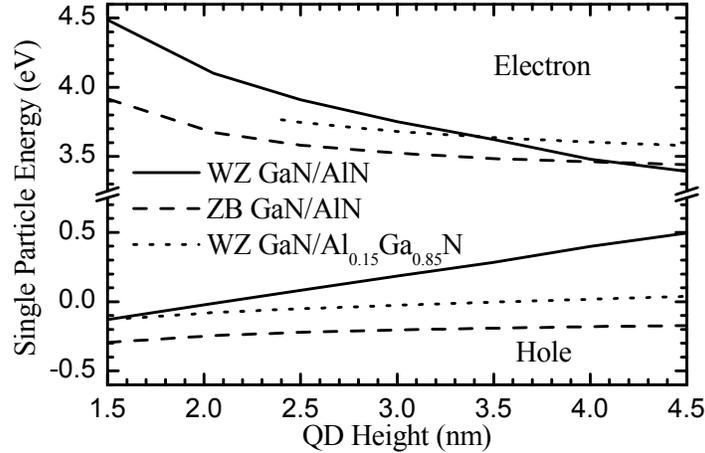

FIG. 7. Electron and hole ground state energy levels as a function of QD height for three kinds of GaN QDs. Electron and hole energies in WZ GaN/Al$_{0.15}$Ga$_{0.85}$N QDs are shown only for those QD heights that allow at least one discrete energy level.

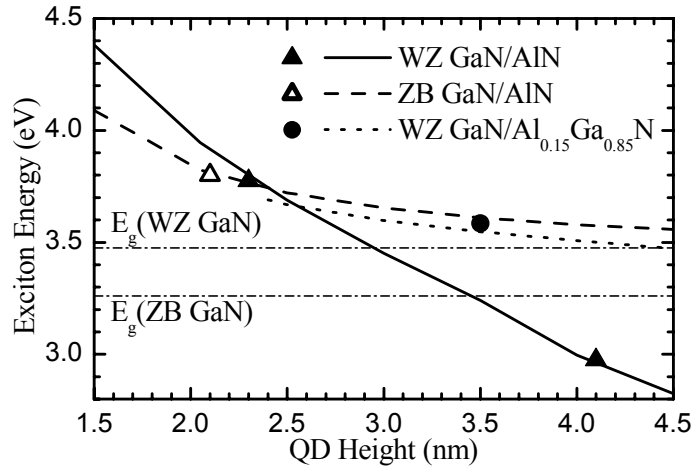

FIG. 8. Exciton ground state energy levels as a function of QD height for three kinds of GaN QDs. Exciton energy in WZ GaN/Al$_{0.15}$Ga$_{0.85}$N QDs is shown only for those QD heights that allow both electron and hole discrete energy levels. Dash-dotted lines indicate bulk energy gaps of WZ GaN and ZB GaN. Filled triangles represent experimental points of Widmann *et al*. [2]; empty triangle is an experimental point of Daudin *et al*. [8]; and filled circle is an experimental point of Ramval *et al*. [4].

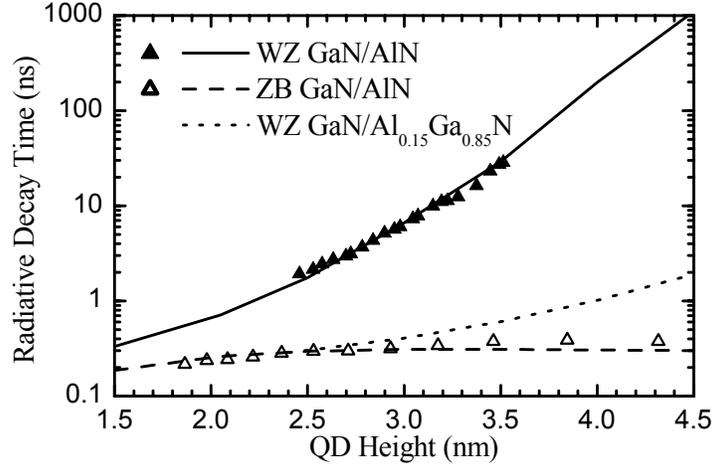

FIG. 9. Radiative decay time as a function of QD height for three kinds of GaN QDs. Radiative decay time in WZ GaN/Al$_{0.15}$Ga$_{0.85}$N QDs is shown only for those QD heights that allow both electron and hole discrete energy levels. Filled and empty triangles represent experimental points from Ref. [24] for WZ GaN/AlN and ZB GaN/AlN QDs, respectively.

## VIII. CONCLUSIONS

We have theoretically investigated the electron, hole, and exciton states, as well as radiative decay times for WZ GaN/AlN, ZB GaN/AlN, and WZ GaN/Al$_{0.15}$Ga$_{0.85}$N quantum dots. Our multi-band model has yielded excitonic energies and radiative decay times that agree very well with available experimental data for all considered GaN QDs. A long radiative decay time in WZ GaN/AlN quantum dots is undesirable for such optoelectronic applications as light-emitting diodes. At the same time, we have shown that at least two other kinds of GaN quantum dots, such as ZB GaN/AlN and WZ GaN/Al$_{0.15}$Ga$_{0.85}$N QDs, have much smaller radiative decay times for the same QD height. It has been also demonstrated that a strong piezoelectric field characteristic for WZ GaN/AlN QDs can be used as an additional tuning parameter for the optical response of such structures. The good agreement of our calculations with the experimental data indicates that our theoretical and numerical models can be applied to study excitonic properties of strained WZ and ZB QD heterostructures.

## ACKNOWLEDGEMENTS

The authors thank Prof. E. P. Pokatilov (State University of Moldova) for many illuminating discussions. This work was supported in part by ONR Young Investigator Award N00014-02-1-0352 to one of the authors (A.A.B.) and the U.S. Civilian Research and Development Foundation (CRDF).

1. F. Widmann, B. Daudin, G. Feuillet, Y. Samson, J. L. Rouviere, and N. Pelekanos, J. Appl. Phys. **83**, 7618 (1998).
2. F. Widmann, J. Simon, B. Daudin, G. Feuillet, J. L. Rouviere, N. T. Pelekanos, and G. Fishman, Phys. Rev. B **58**, R15989 (1998).

3. P. Ramval, S. Tanaka, S. Nomura, P. Riblet, and Y. Aoyagi, Appl. Phys. Lett. **73**, 1104 (1998).
4. P. Ramval, P. Riblet, S. Nomura, Y. Aoyagi, and S. Tanaka, J. Appl. Phys. **87**, 3883 (2000).
5. V. J. Leppert, C. J. Zhang, H. W. H. Lee, I. M. Kennedy, and S. H. Risbud, Appl. Phys. Lett. **72**, 3035 (1998).
6. E. Borsella, M. A. Garcia, G. Mattei, C. Maurizio, P. Mazzoldi, E. Cattaruzza, F. Gonella, G. Battagin, A. Quaranta, and F. D'Acapito, J. Appl. Phys. **90**, 4467 (2001).
7. E. Martinez-Guerrero, C. Adelmann, F. Chabuel, J. Simon, N. T. Pelekanos, G. Mula, B. Daudin, G. Feuillet, and H. Mariette, Appl. Phys. Lett. **77**, 809 (2000).
8. B. Daudin, G. Feuillet, H. Mariette, G. Mula, N. Pelekanos, E. Molva, J. L. Rouviere, C. Adelmann, E. Martinez-Guerrero, J. Barjon, F. Chabuel, B. Bataillou, and J. Simon, Jpn. J. Appl. Phys. **40**, 1892 (2001).
9. A. D. Andreev and E. P. O'Reilly, Phys. Rev. B **62**, 15851 (2000).
10. V. A. Fonoberov, E. P. Pokatilov, and A. A. Balandin, J. Nanosci. Nanotech. **3**, 253 (2003).
11. L. D. Landau and E. M. Lifshitz, *Theory of Elasticity*, 3$^{rd}$ ed. (Pergamon, Oxford, 1986).
12. I. P. Ipatova, V. G. Malyshkin, and V. A. Shchukin, J. Appl. Phys. **74**, 7198 (1993).
13. I. Vurgaftman, J. R. Meyer, and L. R. Ram-Mohan, J. Appl. Phys. **89**, 5815 (2001).
14. K. Shimada, T. Sota, and K. Suzuki, J. Appl. Phys. **84**, 4951 (1998).
15. M. E. Levinshtein, S. L. Rumyantsev, and M. S. Shur, "Properties of Advanced Semiconductor Materials: GaN, AlN, InN, BN, SiC, and SiGe", John Wiley and Sons, New York (2001).
16. E. Ejder, Phys. Stat. Sol. (a) **6**, 445 (1971).
17. F. Bernardini, V. Fiorentini, and D. Vanderbilt, Phys. Rev. B **56**, R10024 (1997).
18. S. M. Komirenko, K. W. Kim, M. A. Stroscio, and M. Dutta, Phys. Rev. B **59**, 5013 (1999).
19. E. P. Pokatilov, V. A. Fonoberov, V. M. Fomin, and J. T. Devreese, Phys. Rev. B **64**, 245328 (2001).
20. F. Mireles and S. E. Ulloa, Phys. Rev. B **62**, 2562 (2000).
21. C. Pryor, M. E. Pistol, and L. Samuelson, Phys. Rev. B **56**, 10404 (1997).
22. V. A. Fonoberov, E. P. Pokatilov, and A. A. Balandin, Phys. Rev. B **66**, 085310 (2002).
23. G. W. Hooft, W. A. Poel, L. W. Molenkamp, and C. T. Foxon, Phys. Rev. B **35**, 8281 (1987).
24. L. S. Dang, G. Fishman, H. Mariette, C. Adelmann, E. Martinez, J. Simon, B. Daudin, E. Monroy, N. Pelekanos, J. L. Rouviere, and Y. H. Cho, J. Kor. Phys. Soc. **42**, S657 (2003); J. Simon, N. T. Pelekanos, C. Adelmann, E. Martinez-Guerrero, R. Andre, B. Daudin, Le Si Dang, and H. Mariette, Phys. Rev. B **68**, 035312 (2003).